\documentclass[cup6b,epsf]{cupbook}

\usepackage{epsf}

\begin{document}

\author[V. Bromm and A. Loeb]{VOLKER BROMM$^1$ and ABRAHAM
LOEB$^2$\\ (1) Astronomy Department, The University of Texas at
Austin, USA\\ (2) Astronomy Department, Harvard University, Cambridge,
USA}

\chapter{GRB Cosmology}

\section*{Abstract}
\noindent Current observations are about to open up a direct window into
the final frontier of cosmology: the first billion years in cosmic history
when the first stars and galaxies formed. Even before the launch of the
{\it James Webb Space Telescope}, it might be possible to utilize Gamma-ray
Bursts (GRBs) as unique probes of cosmic star formation and the state of
the intergalactic medium (IGM) up to redshifts of several tens, when the
first (Population~III) stars had formed. The {\it Swift} mission, or future
satellites such as EXIST, might be the first observatories to detect
individual Population~III stars, provided that massive metal-free stars
were able to trigger GRBs. Spectroscopic follow-up observations of the GRB
afterglow emission would allow to probe the ionization state and metal
enrichment of the IGM as a function of redshift.

\section{Introduction}

One of the important goals in modern cosmology is to understand how the
first stars formed at the end of the cosmic dark ages, and how they
transformed the initially simple, homogeneous universe into a state of ever
increasing complexity (e.g., Barkana \& Loeb 2001; Miralda-Escud\'{e} 2003;
Bromm \& Larson 2004; Ciardi \& Ferrara 2005; Loeb 2006).  The first stars
[so-called Population~III (Pop~III)] are predicted, based on results from
numerical simulations, to have started forming at redshifts $z > 20$, and
to have been predominantly very massive with $M_{\ast} > 100 M_{\odot}$
(e.g., Bromm, Coppi, \& Larson 1999, 2002; Abel, Bryan, \& Norman 2000,
2002; Nakamura \& Umemura 2001; Bromm \& Loeb 2004; Yoshida et al. 2006;
Gao et al. 2007; O'Shea \& Norman 2007).  They had likely played a crucial
role in driving early cosmic evolution by producing ionizing photons and
heavy elements. The initial stages in the reionization of the intergalactic
medium (IGM) have recently been investigated in great detail with one and
three dimensional simulations, showing the expansion of individual H~II
regions around the first stars (Kitayama et al. 2004; Whalen, Abel, \&
Norman 2004; Alvarez, Bromm, \& Shapiro 2006; Johnson, Greif, \& Bromm
2007).  In addition, the first stars were responsible for the initial metal
enrichment of the IGM, because the first supernova (SN) explosions rapidly
dispersed the heavy elements that were produced during the short (several
Myr) lifetime of Pop~III stars into the environment. This process initiated
the extended nucleosynthetic build-up of elements, starting from a
hydrogen/helium mixture to a fully metal-rich gas (e.g., Madau, Ferrara, \&
Rees 2001; Mori, Ferrara, \& Madau 2002; Scannapieco, Ferrara, \& Madau
2002; Thacker, Scannapieco, \& Davis 2002; Bromm, Yoshida, \& Hernquist
2003; Furlanetto \& Loeb 2003, 2005; Wada \& Venkatesan 2003; Daigne et
al. 2004, 2006a; Norman, O'Shea, \& Paschos 2004; Yoshida, Bromm, \&
Hernquist 2004; Greif et al. 2007). 

The early metal enrichment had important implications. In particular, the
character of star formation changed from an early, high-mass dominated
(Pop~III) mode to the familiar, lower-mass (Pop~II) mode, once a threshold
level of enrichment had been reached, the so-called ``{\it critical
metallicity}'', $Z_{\rm crit}\sim 10^{-3.5}Z_{\odot}$ (e.g., Omukai 2000;
Bromm et al. 2001; Bromm \& Loeb 2003a; Schneider et al. 2003, 2006;
Frebel, Johnson, \& Bromm 2007; Smith \& Sigurdsson 2007). It is therefore
important to explore the topology of early metal enrichment, and find
when particular regions in the universe become supercritical (e.g.,
Schneider et al. 2002; Mackey, Bromm, \& Hernquist 2003; Scannapieco,
Schneider, \& Ferrara 2003; Ricotti \& Ostriker 2004; Furlanetto \& Loeb
2005; Greif \& Bromm 2006; Venkatesan 2006).

Gamma-Ray Bursts (GRBs) provide ideal probes of the formation rate and
environmental impact of stars in the high-redshift universe, including the
reionization and metal enrichment of the IGM. The high luminosities of GRBs
make them detectable out to the edge of the visible universe (e.g., Lamb \&
Reichart 2000; Ciardi \& Loeb 2000; Bromm \& Loeb 2002; Naoz \& Bromberg
2007).  GRBs offer the exciting opportunity to detect the first (Pop~III)
stars, one star at a time.  Although the detailed nature of the central
engine that powers the relativistic jets of GRBs is still unknown, recent
evidence indicates that GRBs trace the formation of massive stars (e.g.,
Totani 1997; Wijers et al. 1998; Blain \& Natarajan 2000; Kulkarni et
al. 2000; Porciani \& Madau 2001; Bloom, Kulkarni, \& Djorgovski 2002;
Mesinger, Perna, \& Haiman 2005; Natarajan et al. 2005; Daigne, Rossi, \&
Mochkovitch 2006b; but see Guetta \& Piran 2007). There is growing evidence
that long-duration GRBs are associated with Type Ib/c SNe (e.g., Hjorth et
al. 2003; Matheson et al. 2003; Stanek et al. 2003; see Woosley \& Bloom
2006 for a recent review).  The popular collapsar model for the central
engine of long-duration GRBs (Woosley 1993; MacFadyen, Woosley, \& Heger
2001 and references therein) involves the collapse of a massive star to a
black hole (BH).  Because of their high characteristic masses, a
significant fraction of Pop~III stars might end their lives as a BH,
potentially leading to large numbers of high-redshift GRBs. In the
hierarchical assembly process of halos which are dominated by cold dark
matter (CDM), the first galaxies should have had lower masses (and lower
stellar luminosities) than their low-redshift counterparts. Consequently,
the characteristic luminosity of galaxies or quasars is expected to decline
with increasing redshift. GRB afterglows, which already produce a peak flux
comparable to that of quasars or starburst galaxies at $z\sim 1-2$, are
therefore expected to outshine any competing source at the highest
redshifts, when the first dwarf galaxies have formed in the universe
(e.g. Ciardi \& Loeb 2000; Gou et al. 2004; Inoue, Omukai, \& Ciardi 2007).

The polarization data from the {\it Wilkinson Microwave Anisotropy Probe}
({\it WMAP}) indicates an optical depth to electron scattering of $\sim
9\pm 3$\% after cosmological recombination (Spergel et al. 2007). This
implies that the first stars must have formed before a redshift $z\sim 10$,
and reionized a substantial fraction of the intergalactic hydrogen by that
time (e.g., Cen 2003; Ciardi, Ferrara, \& White 2003; Somerville \& Livio
2003; Wyithe \& Loeb 2003; Yoshida et al. 2004).  Early reionization can be
achieved with plausible star formation parameters in the standard
$\Lambda$CDM cosmology; in fact, the required optical depth can be achieved
in a variety of very different ionization histories since {\it WMAP} places
only an integral constraint on these histories
(e.g., Haiman \& Holder 2003; Greif \& Bromm 2006). One
would like to probe the full history of reionization in order to
disentangle the properties and formation history of the stars that are
responsible for it. GRB afterglows offer the opportunity to detect stars as
well as to probe the ionization state (Barkana \& Loeb 2004) and metal
enrichment level (Furlanetto \& Loeb 2003) of the intervening IGM out to
redshifts $z>10$. The detection of even a very small number of such  
high-redshift bursts could provide interesting constraints on the small-scale
power spectrum of density fluctuations. Models with reduced small-scale
power, such as warm dark matter (WDM) scenarios, could then be ruled out
(Mesinger et al. 2005).

The potential high-redshift GRBs can be identified through infrared
photometry, based on the Ly$\alpha$ break induced by absorption of their
spectrum at wavelengths below $1.216\, \mu {\rm m}\, [(1+z)/10]$. Follow-up
spectroscopy of high-redshift candidates can then be performed on a
10-meter-class telescope. Recently, the ongoing {\it Swift} mission
(Gehrels et al. 2004) has detected a GRB originating at $z\simeq 6.3$
(e.g., Haislip et al. 2006; Kawai et al. 2006), thus demonstrating the
viability of GRBs as probes of the early universe.

There are four main advantages of GRBs relative to traditional cosmic
sources such as quasars:

\noindent {\it (i)} The GRB afterglow flux at a given observed time lag
after the $\gamma$-ray trigger is not expected to fade significantly with
increasing redshift, since higher redshifts translate to earlier times in
the source frame, during which the afterglow is intrinsically brighter
(Ciardi \& Loeb 2000). For standard afterglow lightcurves and spectra, the
increase in the luminosity distance with redshift is compensated by this
cosmological time-stretching effect. This is illustrated in Figure~15.1.

\begin{figure*}
\begin{center}
 \epsfxsize=10cm \epsfbox{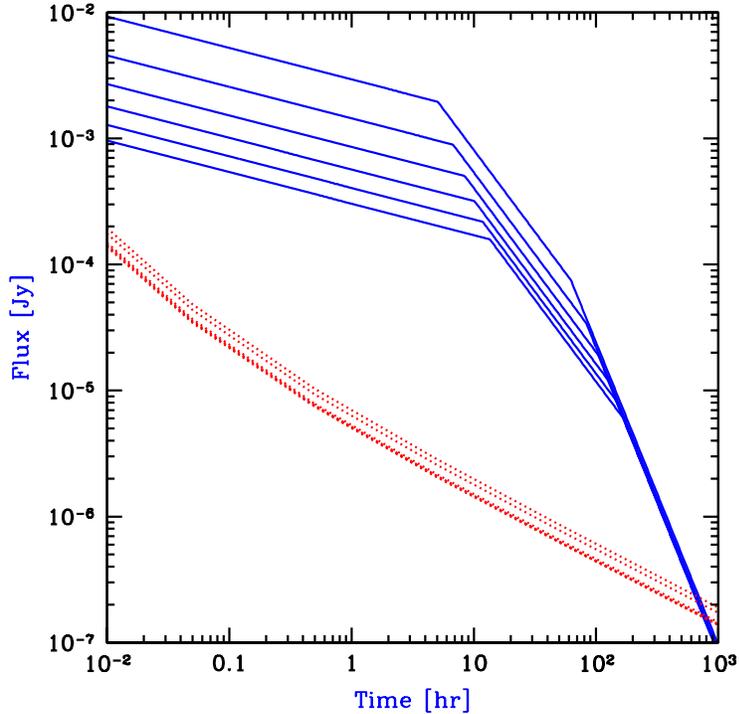}
\end{center}
 \caption{Detectability of high-redshift GRB afterglows as a function
of time since the GRB explosion as measured by the observer (adopted from
Barkana \& Loeb 2004). The GRB afterglow flux (in Jansky) is shown {\it
(solid curves)} at the redshifted Ly$\alpha$ wavelength. Also shown {\it
(dotted curves)} is the detection threshold for {\it JWST} assuming a
spectral resolution $R=5000$ with the near infrared spectrometer, a signal
to noise ratio of 5 per spectral resolution element, and an exposure time
equal to 20\% of the time since the GRB explosion. In each set of curves,
a sequence of redshifts is used, $z=5, 7, 9, 11, 13, 15$, respectively
(from top to bottom).}
 \label{fig1}
\end{figure*}

\noindent {\it (ii)} As already mentioned, in the standard $\Lambda$CDM
cosmology, galaxies form hierarchically, starting from small masses and
increasing their average mass with cosmic time. Hence, the characteristic
mass of quasar black holes and the total stellar mass of a galaxy were
smaller at higher redshifts, making these sources intrinsically fainter
(Wyithe \& Loeb 2002).  However, GRBs are believed to originate from a
stellar mass progenitor and so the intrinsic luminosity of their engine
should not depend on the mass of their host galaxy. GRB afterglows are
therefore expected to outshine their host galaxies by a factor that gets
larger with increasing redshift.

\noindent {\it (iii)} Since the progenitors of GRBs are believed to be
stellar, they likely originate in the most common star-forming galaxies at
a given redshift rather than in the most massive host galaxies, as is the
case for bright quasars (Barkana \& Loeb 2004). Low-mass host galaxies
induce only a weak ionization effect on the surrounding IGM and do not
greatly perturb the Hubble flow around them (e.g., Alvarez et al. 2006;
Johnson et al. 2007).  Hence, the Ly$\alpha$ damping wing should be closer
to the idealized unperturbed IGM case (Miralda-Escud\'{e} 1998) and its
detailed spectral shape should be easier to interpret. Note also that
unlike the case of a quasar, a GRB afterglow can itself ionize at most
$\sim 4\times 10^4 E_{51} M_\odot$ of hydrogen if its UV energy is $E_{51}$
in units of $10^{51}$ ergs (based on the available number of ionizing
photons), and so it should have a negligible cosmic effect on the
surrounding IGM.

\noindent
{\it (iv)} GRB afterglows have smooth (broken power-law) continuum spectra
unlike quasars which show strong spectral features (such as broad emission
lines or the so-called ``blue bump'') that complicate the extraction of IGM
absorption features. In particular, the continuum extrapolation into the
Ly$\alpha$ damping wing (the Gunn-Peterson (1965) absorption trough) during
the epoch of reionization is much more straightforward for the smooth UV
spectra of GRB afterglows than for quasars with an underlying broad
Ly$\alpha$ emission line (Barkana \& Loeb 2004).

In contrast to quasars of comparable brightness, GRB afterglows are
short-lived and release $\sim 10$ orders of magnitude less energy into the
surrounding IGM. Beyond the scale of their host galaxy, GRBs have a
negligible effect on their cosmological environment. Note, however, that
feedback from a single GRB or SN on the gas confined within early dwarf
galaxies could be dramatic, since the binding energy of most galaxies at
$z>10$ is lower than $10^{51}~{\rm ergs}$ (Barkana \& Loeb
2001). Consequently, GRBs are ideal probes of the IGM during the
reionization epoch.  Their rest-frame UV spectra can be used to probe the
ionization state of the IGM through the spectral shape of the Gunn-Peterson
(Ly$\alpha$) absorption trough, or its metal enrichment history through the
intersection of enriched bubbles of SN ejecta from early galaxies
(Furlanetto \& Loeb 2003).  Afterglows that are unusually bright ($>10$mJy)
at radio frequencies could show a detectable forest of 21~cm absorption
lines due to enhanced H~I column densities in sheets, filaments, and
collapsed minihalos within the IGM (Carilli, Gnedin, \& Owen 2002;
Furlanetto \& Loeb 2002; but see Ioka \& M\'{e}sz\'{a}ros 2005).

Another advantage of GRB afterglows is that once they fade away, one may
search for their host galaxies. Hence, GRBs may serve as signposts of the
earliest dwarf galaxies that are otherwise too faint or rare on their own
for a dedicated search to find them. Detection of metal absorption lines
from the host galaxy in the afterglow spectrum, offers an unusual
opportunity to study the physical conditions (temperature, metallicity,
ionization state, and kinematics) in the interstellar medium of these
high-redshift galaxies.  A small fraction ($\sim 10$\%) of the GRB
afterglows are expected to originate at redshifts $z>5$ (Bromm \& Loeb
2002, 2006).  As mentioned, this subset of afterglows can be selected
photometrically using a small telescope, based on the Ly$\alpha$ break at a
wavelength of $1.216\, \mu {\rm m}\, [(1+z)/10]$, caused by intergalactic
H~I absorption.  The challenge in the upcoming years will be to follow-up on
these candidates spectroscopically, using a large (10-meter class)
telescope.  

GRB afterglows are likely to revolutionize observational cosmology and
replace traditional sources like quasars, as probes of the IGM at $z>5$.
The near future promises to be exciting for GRB astronomy as well as for
studies of the high-redshift universe. The {\it Swift} mission (Gehrels et
al. 2004) has already greatly increased the number of GRBs with measured
redshifts, with a mean redshift of $<z>\sim 2$, compared to $<z>\sim 1$ as
estimated for the BATSE sample, including a few bursts detected at
$z>5$.  Future missions, such as the planned EXIST satellite (Grindlay et
al.  2006) which would provide synergy with the near-infrared follow-up
capabilities of {\it JWST}, promise to further advance the emerging
frontier of GRB cosmology.

Here, we will not discuss the possible use of GRBs as standard candles in
probing the expansion history of the universe out to high redshifts, beyond
what is currently possible with Type Ia SNe, to constrain cosmological
parameters (e.g., see Schaefer 2003, 2007). Recently, the utility of GRBs
as standard candles for cosmology has been seriously challenged (e.g.,
Butler et al. 2007; Li 2007), and it is currently not clear whether this
methodology would provide a useful tool for precision cosmology. We will
also not consider short-duration GRBs, as the associated timescale for the
merger of the binary compact-remnant progenitor is typically too long to
probe cosmic evolution during the first hundreds of millions of years after
the big bang.

This chapter is organized as follows. We begin by zooming in on the very small
spatial scales around the GRB progenitor itself (Section 15.2), proceeed to the
intermediate scale of the GRB host system (Section 15.3), and finally
consider the cosmological scale of the surrounding high-$z$ IGM (Section
15.4). We conclude by discussing the utility of GRBs in probing the cosmic
star formation history in the early universe (Section 15.5). This last
section unifies all the different scales, with the ultimate goal of
arriving at a coherent framework for understanding early cosmic star
formation.

\section{Population~III GRB Progenitors}

No GRB from a Pop~III progenitor has been observed to date, but it is
intriguing to ask: {\it What is the expected signature of GRBs that were
triggered by the death of a massive Pop~III star?}  The information that a
particular GRB originated at a high redshift is not sufficient to establish
the case for the nature of its progenitor. For example, the currently
highest-redshift GRB at $z\simeq 6.3$ (GRB~050904) clearly did not
originate from a Pop~III progenitor, given that the inferred level of metal
enrichment in the host system is a few percent of the solar abundance
(e.g., Campana et al. 2007). Pregalactic metal enrichment was 
inhomogeneous, and we expect normal Pop~I and II stars to exist in galaxies
that were already metal-enriched at these high redshifts. Pop~III and
Pop~I/II star formation is thus predicted to have occurred concurrently at
$z > 5$. 

However, it is plausible that the high mass-scale for Pop~III stars is
reflected in the observational signature of the resulting GRBs.
Specifically, we estimate that the mass of the BH at the center of the
collapsar is a factor of 10 larger for Pop~III, and the Pop~III GRB
distribution may thus be biased toward longer-duration events. It remains
an open question as to what the exact dependence of the GRB duration,
$T_{\rm dur}$, is on the collapsar BH mass, but one might guess a simple
linear relation to first order: $T_{\rm dur}\propto R_{\rm Sch}/c \propto
M_{\rm BH}$, where $R_{\rm Sch}$ is the Schwarzschild radius. The same
linear scaling with mass might apply to the total energy released in
$\gamma$ rays, if one assumes that the energy yield is proportional to the
total (rest mass) energy reservoir of the BH. The properties of the
corresponding Pop~III GRB afterglow emission are more uncertain (e.g.,
Ciardi \& Loeb 2000; Gou et al. 2004; Inoue et al. 2007), because they
depend sensitively on the circumburst density in the Pop~III host system
(see the next section).

Empirically, it is estimated that the ratio between the rates of
intermediate-redshift GRBs and core collapse SNe is $\sim 10^{-3}$ (e.g.,
Langer \& Norman 2006). It is then important to estimate the corresponding
GRB frequency for Pop~III stars (e.g., Bromm \& Loeb 2006).
Conservatively, one could simply assume a constant efficiency of forming
GRBs per unit mass of stars, independent of time and stellar
population. Such an assumption, however, could be seriously in error.
There are conflicting trends that could move the frequency of Pop~III GRBs
away from their low-redshift counterparts.  Metal-free stars are predicted
to have been massive (Abel et al. 2002; Bromm et al. 2002) and their
extended envelopes may have suppressed the emergence of relativistic jets
out of their surface (even if such jets were produced through the collapse
of their core to a spinning black hole). On the other hand, low-metallicity
stars are expected to have had weak winds with little angular momentum loss
during their evolution, and so they may have preferentially produced
rotating central configurations that make GRB jets after core
collapse. {\it Should GRBs be a common phenomenon among the first
metal-free stars?}

Long-duration GRBs appear to be associated with Type Ib/c supernovae (e.g.,
Stanek et al. 2003), namely progenitor massive stars that have lost their
hydrogen envelope. The lack of an envelope was anticipated theoretically by
the collapsar model, in which the relativistic jets produced by core
collapse to a black hole are unable to emerge relativistically out of the
stellar surface if the hydrogen envelope is retained (MacFadyen et
al.~2001). The question then arises as to whether the lack of metal-line
opacity that is essential for radiation-driven winds in metal-rich stars
(e.g., Kudritzki \& Puls 2000), would make a Pop~III star retain its
hydrogen envelope, thus quenching any relativistic jets and GRBs. There
are, however, other mechanisms that might allow a Pop~III star to shed its
hydrogen envelope: {\it (i)} violent pulsations, particularly in the mass
range $100$--$140 M_{\odot}$, or {\it (ii)} a wind driven by helium
lines. The outer atmospheres of these unusual stars are in a state where
gravity only marginally exceeds radiation pressure due to
electron-scattering (Thomson) opacity (e.g., Bromm, Kudritzki, \& Loeb
2001). Adding the small, but still non-negligible contribution from the
bound-free opacity provided by singly-ionized helium, may be able to unbind
the atmospheric gas.  Therefore, mass-loss could occur even in the absence
of dust, or any heavy elements.

Assuming that long-duration GRBs are indeed produced by the collapsar
mechanism, it is a matter of active debate whether a massive star with a
close binary companion is the primary route to forming a GRB progenitor
(rapidly spinning black hole) or whether single star progenitors are
required (e.g., Petrovic et al. 2005).  The former was suggested by Bromm
\& Loeb (2006) for Pop~III bursts to reconcile the two competing
requirements of envelope removal in the absence of metal-line opacity and
avoiding angular momentum loss. The same binary model was critically
analysed by Belczynski et al. (2007), who argue that only a small fraction
of possible close Pop~III binaries will retain or acquire the spin required
for the collapsar engine, when angular momentum transport is modelled with
sufficient realism. It will be important, however, to explore more fully
the complex physics of mass and angular momentum loss from Pop~III stars,
and to verify whether traditional stellar evolution models, such as those
adopted by Belczynski et al. (2007), apply to metal-free
stars. Interestingly, there has been suggestions for a single-star GRB
progenitor that could circumvent some of the difficulties with realistic
binary models (e.g., Yoon \& Langer 2005; Woosley \& Heger 2006; Yoon,
Langer, \& Norman 2006). The single-star models invoke stars that
experience little mass loss and undergo chemically homogeneous nuclear
burning, driven by rapid rotation. The star would then never leave the main
sequence and expand into a red giant, so that a relativistic jet could
pierce through the stellar material without being quenched. The requirement
for reduced mass loss leads in turn to low metallicities ($<0.1 Z_{\odot}$)
in the stellar atmospheres. Such a possible low-metallicity bias would be
important in deriving the GRB redshift distribution from the cosmic star
formation history (see \S 15.5).

\section{Physical Properties of GRB Hosts}

In order to predict the observational signature of high-redshift GRBs, we
need to examine the unusual circumburst environment inside the systems
that hosted the first stars. In particular, the properties of the afterglow
emission that is associated with Pop~III GRBs will sensitively depend on
the density structure in the Pop~III host system (e.g., Ciardi \& Loeb
2000; Gou et al. 2004; Inoue et al. 2007).  In general, higher densities
translate to larger fluxes in the observer-frame near-infrared waveband,
whereas the X-ray afterglows show almost no dependence on circumburst
density (Gou et al. 2004).

{\it What do we know about the host systems of Pop~III star formation, and
therefore of high-redshift GRBs?}  Within variants of the popular CDM model
for structure formation, where small objects form first and subsequently
merge to build up more massive ones, the first stars are predicted to form
in the redshift range of $z\sim 20$--$30$ (Tegmark et al. 1997; Barkana \&
Loeb 2001; Yoshida et al. 2003).  Figure~15.2 shows results from a
cosmological simulation, illustrating the projected density field at
$z=20$. The bright knots at the intersections of the filamentary network
are so-called `minihalos' of total mass (dark matter plus gas) $\sim 10^6
M_{\odot}$. These objects are the sites for the formation of the first
stars, and thus are the potential hosts of the highest-redshift GRBs.
\begin{figure}[t]
\vspace{0.5cm}
\begin{center}
 \epsfxsize=9.5cm \epsfbox{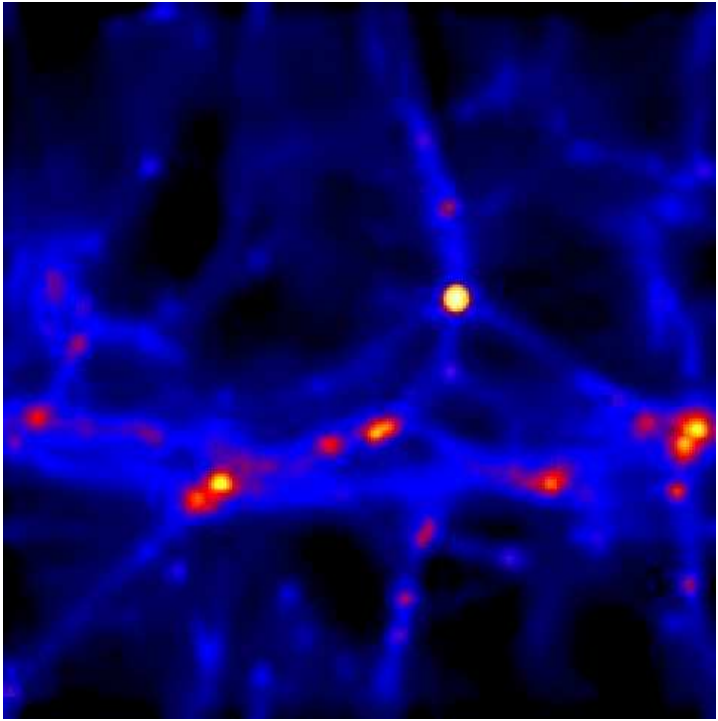}
\end{center}
\vspace{0cm}
\caption{Density field within a standard $\Lambda$CDM cosmology at $z=20$.
The box has a physical size of $\sim 10$~kpc. Shown is the projected gas
density (with darker shades corresponding to lower density) which closely
follows the dynamically dominant dark matter component.
The bright knots at the
intersection of the filamentary network are the sites where the first stars
formed, and where the first GRBs might have exploded (from Bromm et
al. 2003).}
\label{fig2}
\end{figure}

The identification of the environments in which potential Pop~III GRBs and
their afterglows occur, provides the motivation for studying the gas at the
center of their host minihalos just before a massive star ends its life and
the GRB explosion is triggered.  This problem breaks down into two related
questions: {\it (i)} what type of stars (in terms of mass, rotation, and
clustering properties) will form in each minihalo?, and {\it (ii)} how will
the ionizing radiation from each star modify the density structure of the
surrounding gas? These two questions are fundamentally intertwined. The
production of ionizing photons depends strongly on the stellar mass, which
in turn is determined by how the accretion flow onto the growing protostar
proceeds under the influence of this radiation field. In other words, the
assembly of the Pop~III stars and the development of an H~II region around
them proceed simultaneously, and affect each other. The shallow potential
wells in the host minihalos, with corresponding circular velocities of a
few km~s$^{-1}$, are unable to retain photo-ionized gas, so that the gas is
effectively blown out of the minihalo. The resulting photo-evaporation has
been studied with 3D radiative transfer calculations (e.g., Alvarez et
al. 2006), where one massive Pop~III star at the center of the minihalo
acts as an embedded point source, that also take into account the
hydrodynamic response of the photo-heated gas.  It is possible to
understand the key physics of the photo-evaporation from minihalos with the
self-similar solution for a champagne flow (Shu et al. 2002).  In Figure~15.3,
we show the self-similar density evolution, scaled with parameters
appropriate for the Pop~III case.
\begin{figure}[t]
\begin{center}
 \epsfxsize=10cm \epsfbox{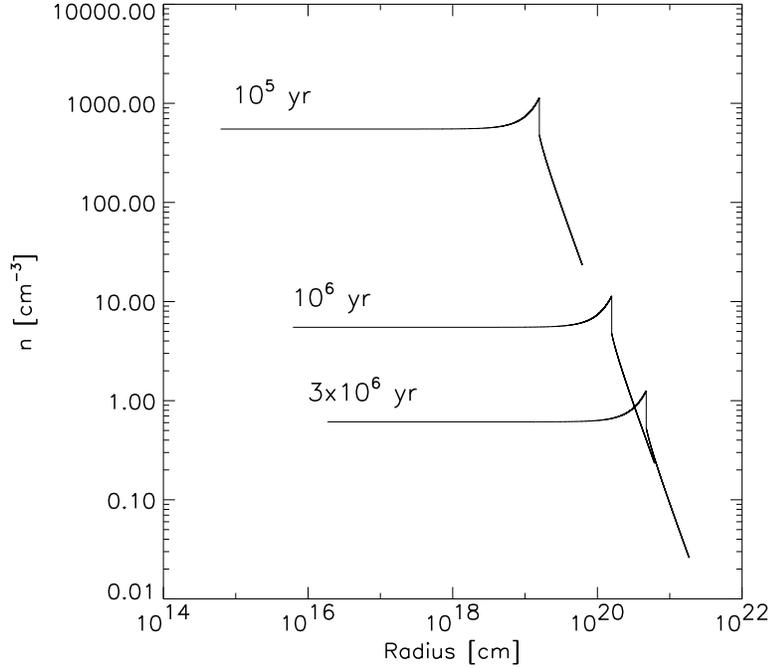}
\end{center}
\vspace{-0.5cm}
\caption{Effect of photoheating from a Population~III star on the density
profile in a high-redshift minihalo.  The curves, labeled by the time after
the onset of the central point source, are calculated according to a
self-similar model for the expansion of an H~II region. Numerical
simulations closely conform to this analytical behavior. Note that the
central density is significantly reduced by the end of the life of a
massive star, and that a central core has developed with a nearly constant
density.}
\label{fig3}
\end{figure}
Note that the central density is significantly reduced by the end of the
life of a massive star, and that a central core has developed in which the
density is nearly constant. Such a flat density profile is markedly
different from that created by stellar winds ($\rho \propto
r^{-2}$). Winds, and consequently mass-loss, may not be important for
massive Pop~III stars (e.g., Baraffe, Heger, \& Woosley 2001; Kudritzki
2002; but see 15.2), and such a flat density profile may be characteristic
of GRBs that originate from metal-free Pop~III progenitors.

\begin{figure}[t]
\vspace{0.5cm}
\begin{center}
 \epsfxsize=12cm \epsfbox{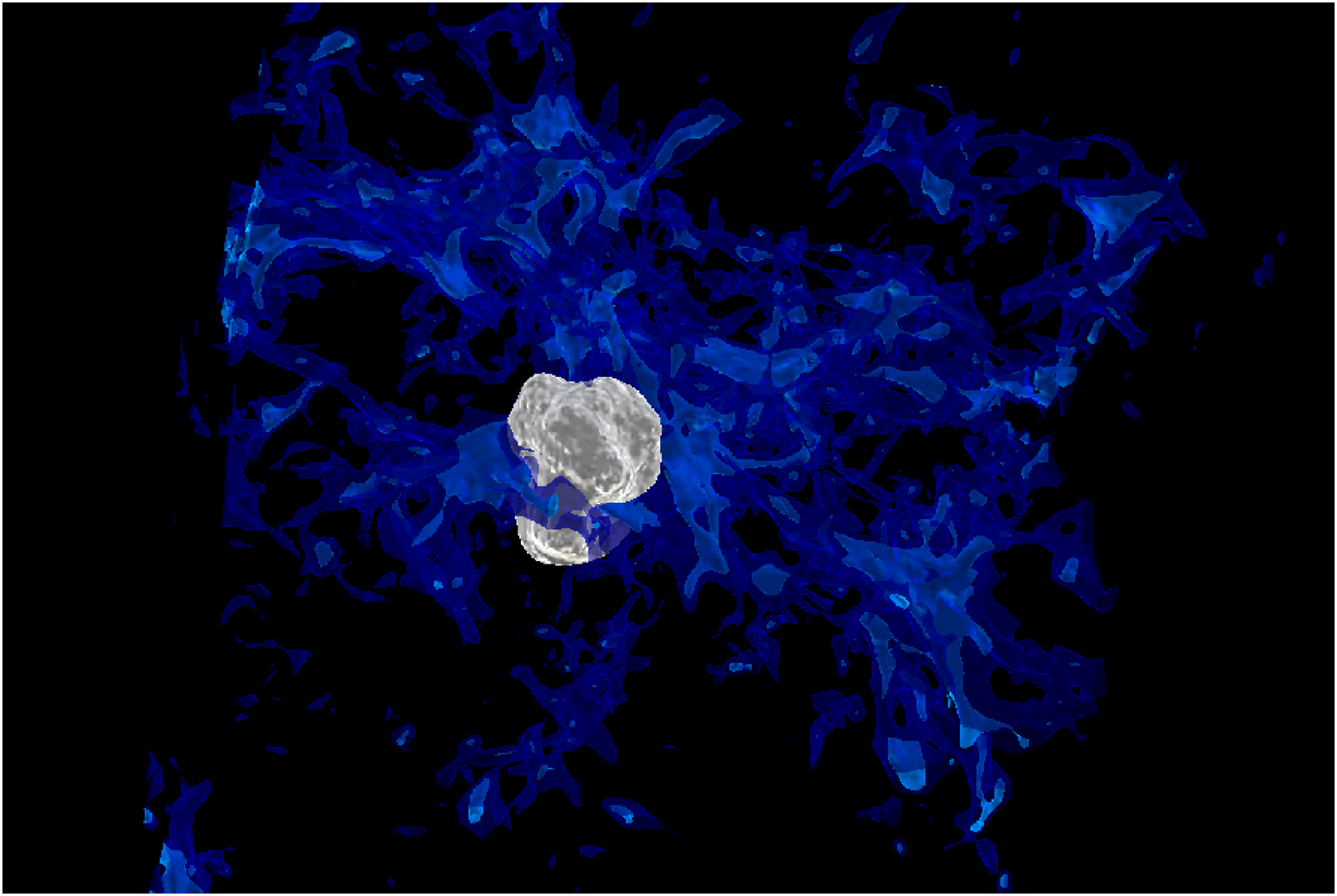}
\end{center}
\vspace{0cm}
\caption{Radiative feedback from the first stars (adopted from Johnson et
al. 2007).  The blue-dark contours show the density field at $z\sim 20$
within a cosmological box of physical size $\sim 30$~kpc. At the center of
the box, a single Pop~III star with $100 M_{\odot}$ has formed, creating a
bubble of ionized radiation ({\it white contour}) that reaches a maximum
size of $\sim 5$~kpc (physical).  The radiative feedback is fairly
localized in extent, and leaves much of the surrounding IGM
undisturbed. This snapshot shows the situation that would be present just
before the Pop~III star dies, possibly triggering a GRB explosion in the
process. (Visualization courtesy of Paul Navr\'{a}til at the Texas Advanced
Computing Center.)}
\label{fig4}
\end{figure}

\begin{figure}[t]
\begin{center}
 \epsfxsize=9.5cm \epsfbox{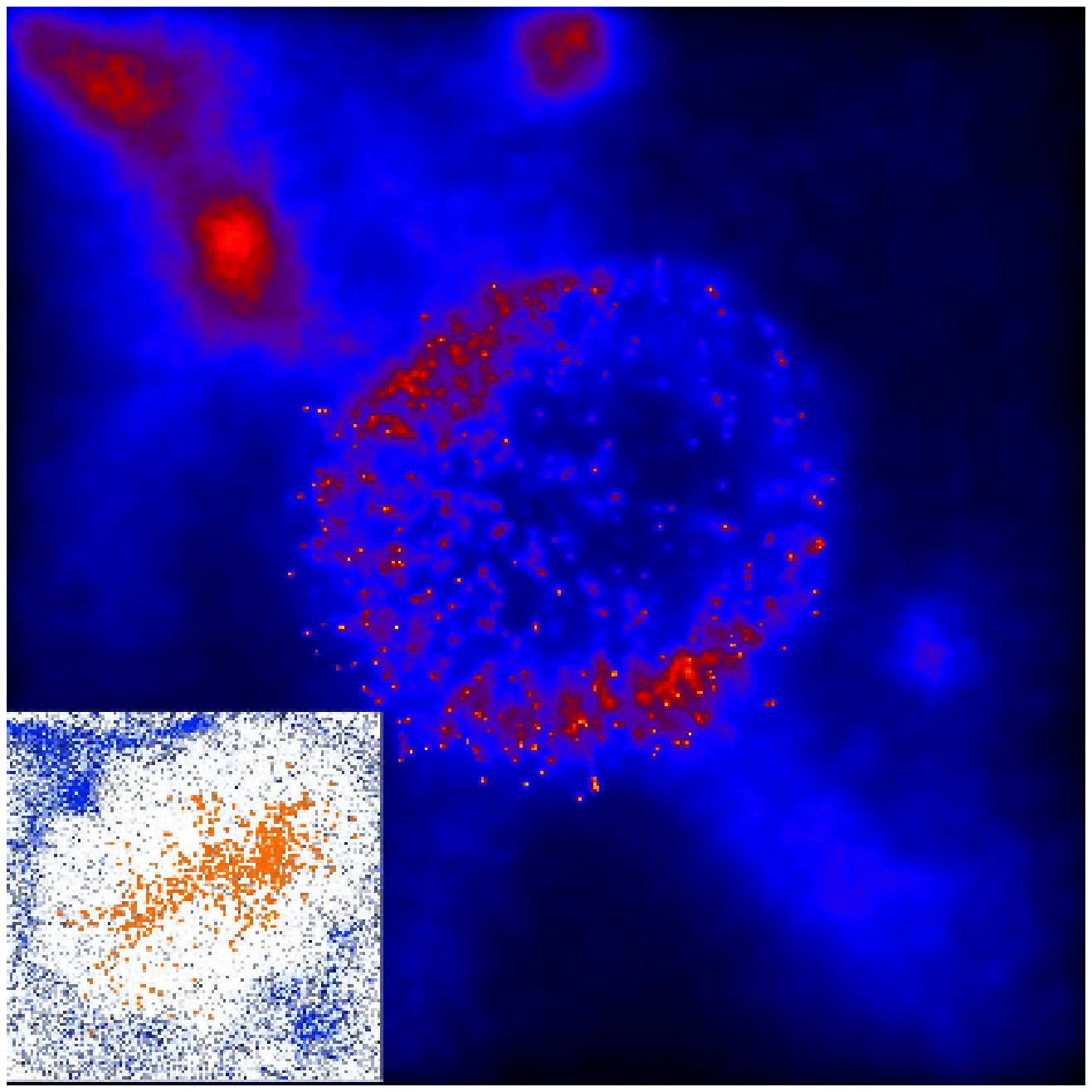}
\end{center}
\vspace{0cm}
\caption{Chemical feedback from the first stars (adopted from Bromm et
al. 2003).  SN explosion in the high-redshift universe that ends the life
of a $200 M_{\odot}$ Pop~III star. The snapshot is taken $\sim 10^6$~yr
after the explosion with total energy $E_{\rm SN}\simeq 10^{53}$~ergs. We
show the projected gas density within a box of linear size 1~kpc. The SN
bubble has expanded to a radius of $\sim 200$~pc, having evacuated most of
the gas in the minihalo. {\it Inset:} Distribution of metals. The stellar
ejecta ({\it red dots}) trace the metals and are embedded in pristine
metal-poor gas ({\it blue dots}).  }
\label{fig5}
\end{figure}

The ultimate modeling goal is to carry out self-consistent simulations of
the protostellar accretion process, taking account of the radiative
feedback on its dynamics. The required coupled radiation hydrodynamics
simulations are just beyond the cutting-edge of existing numerical
capabilities.  In the meantime, we can gain first insight by considering
the results from a somewhat idealized 3D numerical simulation of the
protostellar accretion flow inside a minihalo at $z\sim 20$ (Bromm \& Loeb
2004).  In this simulation, one high-density clump has formed at the center
of the minihalo, possessing a gas mass of a few hundred solar masses. Soon
after its formation, the clump becomes gravitationally unstable and
undergoes runaway collapse. Once the gas has exceeded a threshold density
of $10^{7}$ cm$^{-3}$, a sink particle is inserted into the simulation.
This choice for the density threshold ensures that the local Jeans mass is
resolved throughout the simulation.  The clump (i.e., sink particle) has an
initial mass of $M_{\rm Cl}\simeq 200M_{\odot}$, and grows subsequently by
ongoing accretion of surrounding gas. High-density clumps with such masses
result from the chemistry and cooling rate of molecular hydrogen, H$_{2}$,
which imprint characteristic values of temperature, $T\sim 200$~K, and
density, $n\sim 10^{4}$ cm$^{-3}$, into the metal-free gas (see Bromm et
al.~2002).  Evaluating the Jeans mass for these characteristic values gives
$M_{J}\sim \mbox{\ a few \ }\times 10^{2}M_{\odot}$, which is close to the
initial clump mass found in the simulation. The central clump is clearly
not a star yet.  To probe the subsequent fate of the clump, Bromm \& Loeb
(2004) have refined the resolution in the clump region to follow the
collapse to higher densities (see Bromm \& Loeb 2003b for a description of
the refinement technique). The refined simulation allows one to study the
three-dimensional accretion flow around the protostar (see also Omukai \&
Palla 2001, 2003; Ripamonti et al. 2002; Tan \& McKee 2004). The gas now
reaches densities of $10^{12}$ cm$^{-3}$ before being incorporated into a
central sink particle. At these high densities, three-body reactions
(Palla, Salpeter, \& Stahler 1983) have converted the gas into a fully
molecular form. The growth of this molecular core is followed for the first
$\sim 10^{4}$~yr after its formation, making the idealized assumption that
the protostellar radiation does not affect the accretion flow.  The
accretion rate is initially very high, $\dot{M}_{\rm acc}\sim 0.1
M_{\odot}$~yr$^{-1}$, and subsequently declines roughly as a power law of
time. The mass of the molecular core, taken as a crude estimate for the
protostellar mass, grows with time $t$ approximately as: $M_{\ast}\sim \int
\dot{M}_{\rm acc}{\rm d}t \simeq 0.8 M_{\odot}(t/1\mbox{\ yr})^{0.45}$.  A
robust upper limit for the final mass of the star is then:
$M_{\ast}(t=3\times 10^{6}{\rm yr})\sim 500 M_{\odot}$. In deriving this
upper bound, we conservatively assumed that accretion cannot go on for
longer than the total lifetime of a massive star of a few Myr.  The
numerical results can be understood within the general theoretical
framework of how stars form (see Larson 2003).  Star formation typically
proceeds from the `inside-out', through the accretion of gas onto a central
hydrostatic core.  Whereas the initial mass of the hydrostatic core is very
similar for primordial and present-day star formation (Omukai \& Nishi
1998), the accretion process -- ultimately responsible for setting the
final stellar mass -- is expected to be rather different. On dimensional
grounds, the accretion rate is simply related to the sound speed ($c_s$)
cubed over Newton's constant (or equivalently given by the ratio of the
Jeans mass and the free-fall time): $\dot{M}_{\rm acc}\sim c_s^3/G \propto
T^{3/2}$. A simple comparison of the temperatures in present-day star
forming regions (with a temperature $T\sim 10$~K) with those in primordial
ones ($T\sim 200-300$~K) already indicates a difference in the accretion
rate of more than two orders of magnitude.

{\it Can a Population~III star ever reach this asymptotic mass limit?}  The
answer to this question is not yet known with any certainty, and it depends
on whether the accretion from a dust-free envelope is eventually terminated
by feedback from the star (e.g., Omukai \& Palla 2001, 2003; Ripamonti et
al. 2002; Omukai \& Inutsuka 2002; Tan \& McKee 2004).  The standard
mechanism by which accretion may be terminated in metal-rich gas, namely
radiation pressure on dust grains (Wolfire \& Cassinelli 1987), is
obviously not effective for gas with a primordial composition. Recently, it
has been speculated that accretion could instead be turned off through the
formation of an H~II region (Omukai \& Inutsuka 2002), or through the
radiation pressure exerted by trapped Ly$\alpha$ photons (Tan \& McKee
2004). The termination of the accretion process defines the current
unsolved frontier in studies of Pop~III star formation. Understanding
the GRB circumburst density structure at the highest redshifts 
requires further progress in the Pop~III protostellar accretion problem.

The first galaxies may be surrounded by a shell of highly enriched material
that was carried out in a SN-driven wind. A GRB in that galaxy may show
strong absorption lines at a velocity separation associated with the wind
velocity.  Modelling these winds allows one to calculate the absorption
profile in the featureless spectrum of a GRB afterglow. This will allow 
using the observed spectra of high-$z$ GRBs to directly probe the degree
of metal enrichment in the vicinity of the first star forming regions (see
Furlanetto \& Loeb 2003 for a semi-analytic treatment).  As the early
afterglow radiation propagates through the interstellar environment of the
GRB, it will likely modify the gas properties close to the source; these
changes could in turn be noticed as time-dependent spectral features in the
spectrum of the afterglow and used to derive the properties of the gas
cloud (density, metal abundance, and size). Perna \& Loeb (1998) showed
that the UV afterglow radiation can induce detectable changes to the
interstellar absorption features of the host galaxy; Waxman \& Draine
(2000) and Fruchter, Krolik, \& Rhoads (2001) examined the destruction of
dust by the GRB X-rays, and Draine \& Hao (2002) considered the destruction
of molecules near the GRB source.  Quantitatively, all of the effects
mentioned above strongly depend on the exact properties of the gas in the
host system.

\section{Probing the High-Redshift IGM}

The first stars transformed the early universe by ionizing the cosmic gas
and enriching it with heavy elements that were not produced in the big
bang. Understanding this cosmic metamorphosis is a key driver in modern
cosmology. The goal is to complete the missing pages in our photo album of
the universe which started at $z\sim 10^3$ when the universe became
transparent and ended at the present epoch (Loeb 2006). Great progress has
been made so far in studying the emergence of cosmic structure with
numerical simulations. In Figures~15.4 and 15.5, we show simulations of the
radiative and chemical feedback from the first star to form in a
cosmological simulation box. The radiative feedback represents the initial
stage in the protracted process of reionization, whereas the chemical
feedback is responsible for dispersing the first heavy elements into the
chemically pristine IGM. Despite the importance for cosmology, we currently
have only few and rather indirect observational constraints that would
allow us to test our numerical and theoretical modelling of structure
formation at high-redshift.  GRBs offer the exciting prospect of
directly probing the physical conditions in the IGM during the first
billion years, when the first stars, galaxies and quasars emerged.

\begin{figure}[t]
\begin{center}
 \epsfxsize=9.5cm \epsfbox{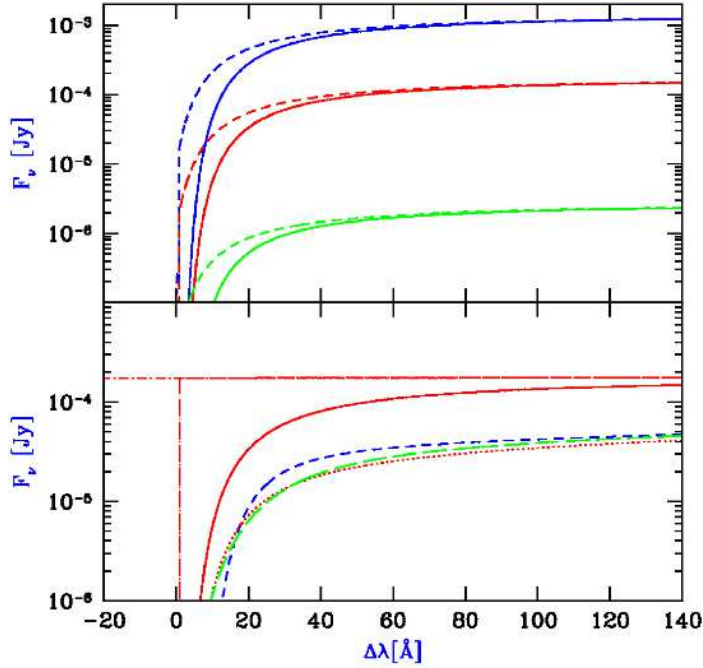}
\end{center}
\vspace{0cm}
\caption{Probing the ionization state of the IGM (from Barkana \& Loeb
2004).  IGM absorption profiles in GRB afterglows, presented in terms of
the flux per unit frequency $F_{\nu}$ versus the observed wavelength shift
from the Ly$\alpha$ resonance $\Delta \lambda$. {\it Top:} Predicted
spectrum including H~I absorption from the IGM (both resonant and damping
wing), for host galaxies with source age $10^7$~yr, an escape fraction of
ionizing radiation $f_{\rm esc}=10$\%, and a normal IMF ({\it solid
curves}) or a source age $10^8$~yr, $f_{\rm esc}=90$\%, and a Pop~III IMF
({\it dashed curves}). The observed time after the burst is ({\it top to
bottom}) 1~h, 1~day, and 10~days.  {\it Bottom:} Predicted spectra 1~day
after a GRB for a host galaxy with a source age of $10^7$~yr, $f_{\rm
esc}=10$\%, and a normal IMF. Shown is the unabsorbed GRB afterglow ({\it
dot--short--dashed curve}; essentially horizontal), the afterglow with
resonant IGM absorption only ({\it dot--long--dashed curve}), and the
afterglow with full (resonant and damping wing) IGM absorption ({\it solid
curve}).  Also shown, with 1.7~mag of extinction, are the afterglow with
full IGM absorption ({\it dotted curve}) and attempts to fit this profile
with a damped Ly$\alpha$ absorption system in the host galaxy ({\it dashed
curves}).}
\label{fig6}
\end{figure}

GRB afterglow emission at wavelengths close to the Ly$\alpha$ resonance
could potentially provide a sensitive diagnostic of the ionization fraction
in the IGM surrounding the explosion site (e.g., Miralda-Escud\'{e} 1998;
Barkana \& Loeb 2004). Specifically, the red damping wing with its reduced
cross section for absorption compared to the line resonance allows to
measure the IGM neutral fraction with high precision. In Figure~15.6, we show
an illustrative example of this technique. High-resolution spectroscopy
around the Ly$\alpha$ resonance in the afterglow spectrum of GRB~050904 at
$z=6.3$ (Totani et al. 2006) and of GRB~060927 at $z=5.47$ (Ruiz-Velasco et
al. 2007) revealed that the red damping wing is already saturated due to a
large column density of neutral hydrogen in the host galaxy ($\log N_{\rm
HI}> 21.6$~cm$^{-2}$), so that the spectra do not allow any inferences on
the IGM neutral fraction at those redshifts. However, it is expected in
hierarchical models of structure formation that the masses of star forming
systems would decrease with increasing redshift, so that the neutral column
densities will become smaller at higher $z$. In such a case, the damping
wing will eventually reflect the conditions in the true IGM and not in the
host systems.

\begin{figure}[t]
\begin{center}
 \epsfxsize=9.5cm \epsfbox{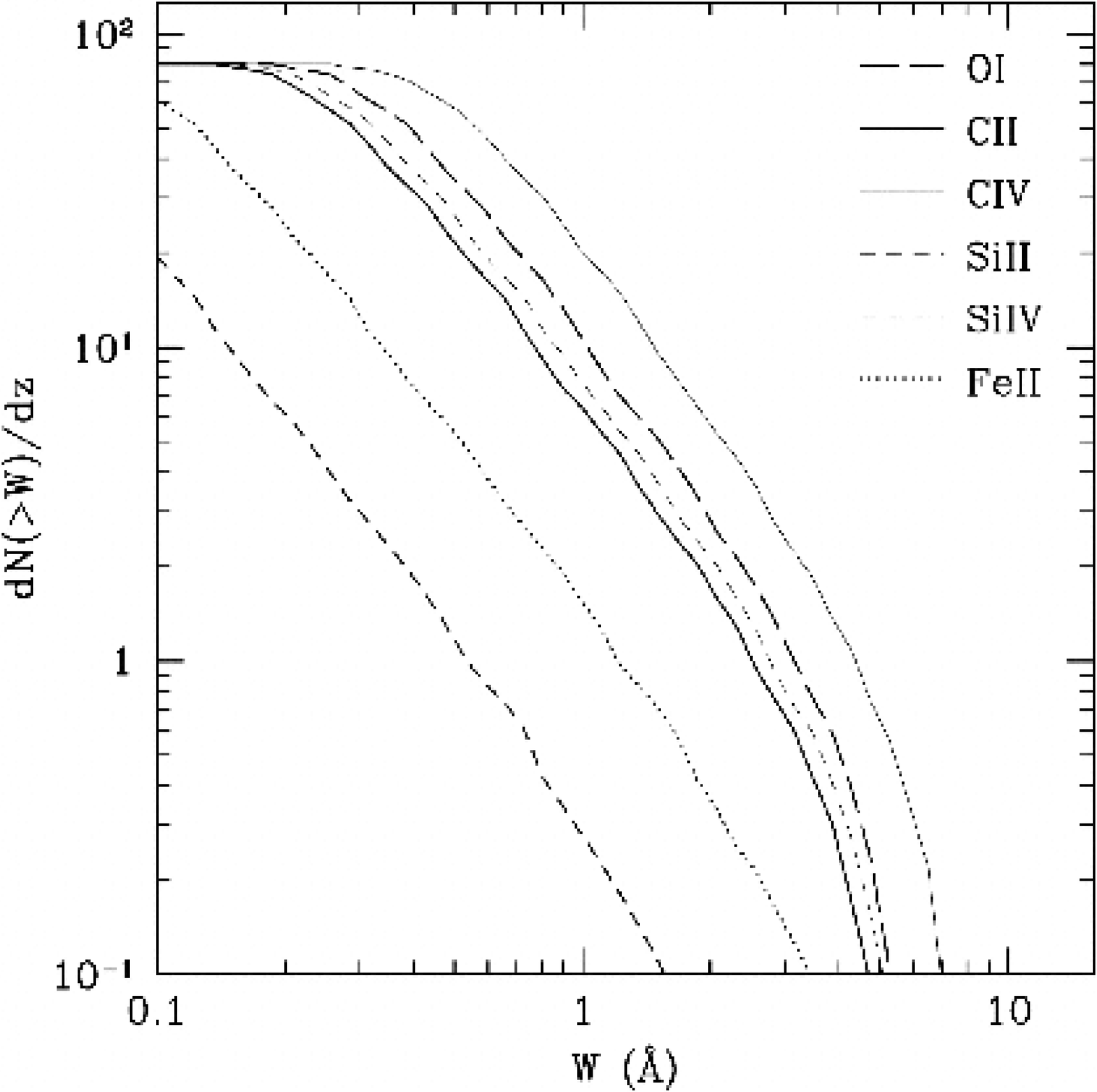}
\end{center}
\vspace{0cm}
\caption{Theoretical prediction for the feasibility probing the metal
enrichment of the IGM at high redshifts (from Furlanetto \& Loeb 2003).
The different curves indicate the expected number of intersections by
absorbers with an equivalent width above $W$ (in \AA ) per unit redshift at
$z=8$, along the line-of-sight to a point source at a redshift $z>8$. The
different lines correspond to specific ionic species, as indicated in the
figure.  }
\label{fig7}
\end{figure}

In addition, high-redshift GRB afterglows also provide ideal `backlights'
to probe the IGM metal enrichment at different times. We illustrate this in
Figure~15.7 (see Furlanetto \& Loeb 2003 for additional details).

\section{Cosmic Star Formation at High Redshifts}

\begin{figure}[t]
\begin{center}
 \epsfxsize=9.5cm \epsfbox{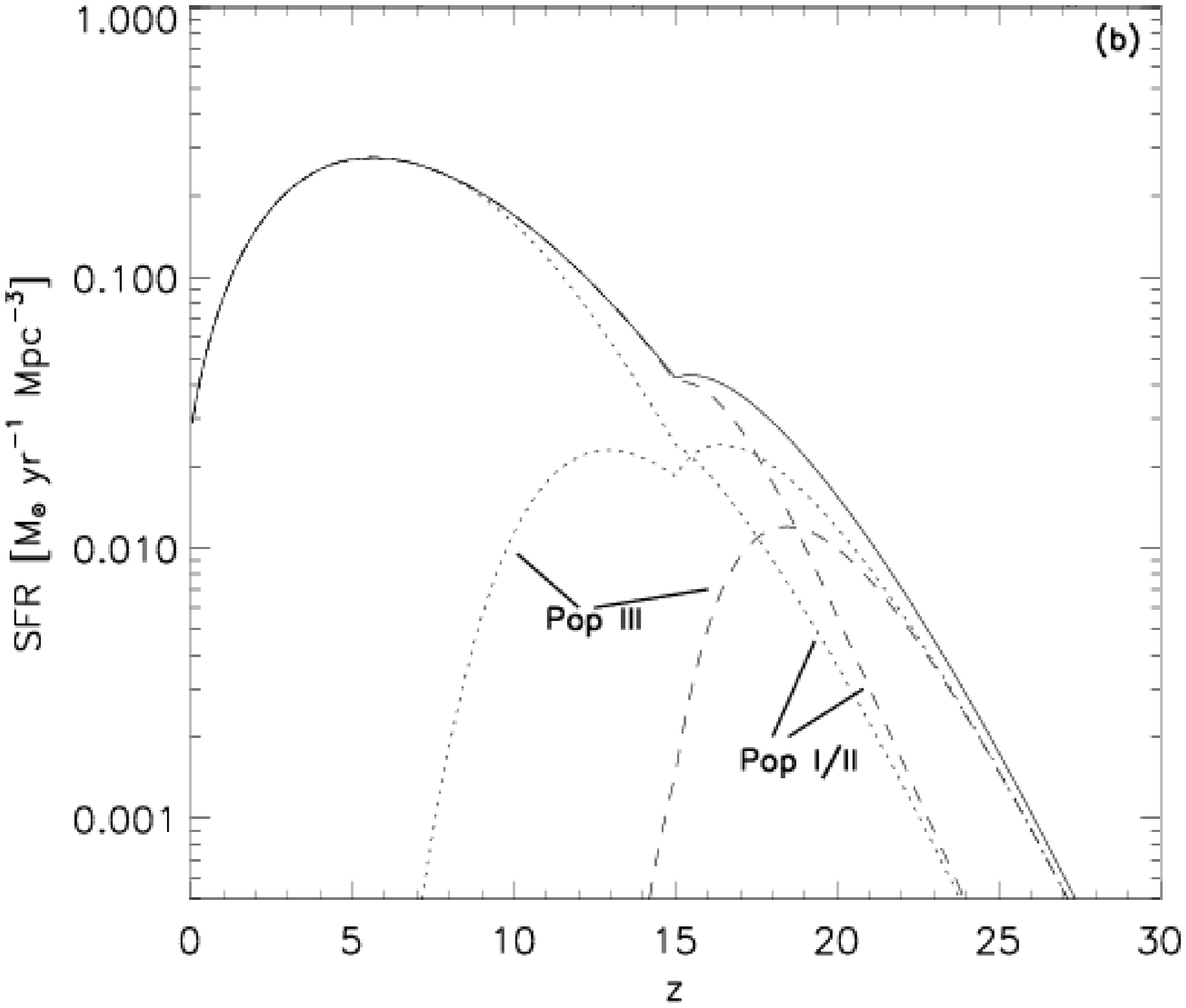}
\end{center}
\vspace{-0.5cm}
\caption{Cosmic star formation rate (SFR) in units of
$M_{\odot}$~yr$^{-1}$~(comoving Mpc)$^{-3}$, as a function of redshift
(from Bromm \& Loeb 2006).  We assume that cooling in primordial gas is due
to atomic hydrogen only, a star formation efficiency of $\eta_\ast=10\%$,
and reionization beginning at $z_{\rm reion}\approx 17$.  {\it Solid line:}
Total comoving SFR.  {\it Dotted lines:} Contribution to the total SFR from
Pop~I/II and Pop~III for the case of weak chemical feedback.  {\it Dashed
lines:} Contribution to the total SFR from Pop~I/II and Pop~III for the
case of strong chemical feedback.  Pop~III star formation is restricted to
high redshifts, but extends over a significant range, $\Delta z\sim
10$--$15$.
\label{fig8}
}
\end{figure}

\begin{figure}[t]
\begin{center}
 \epsfxsize=9.5cm \epsfbox{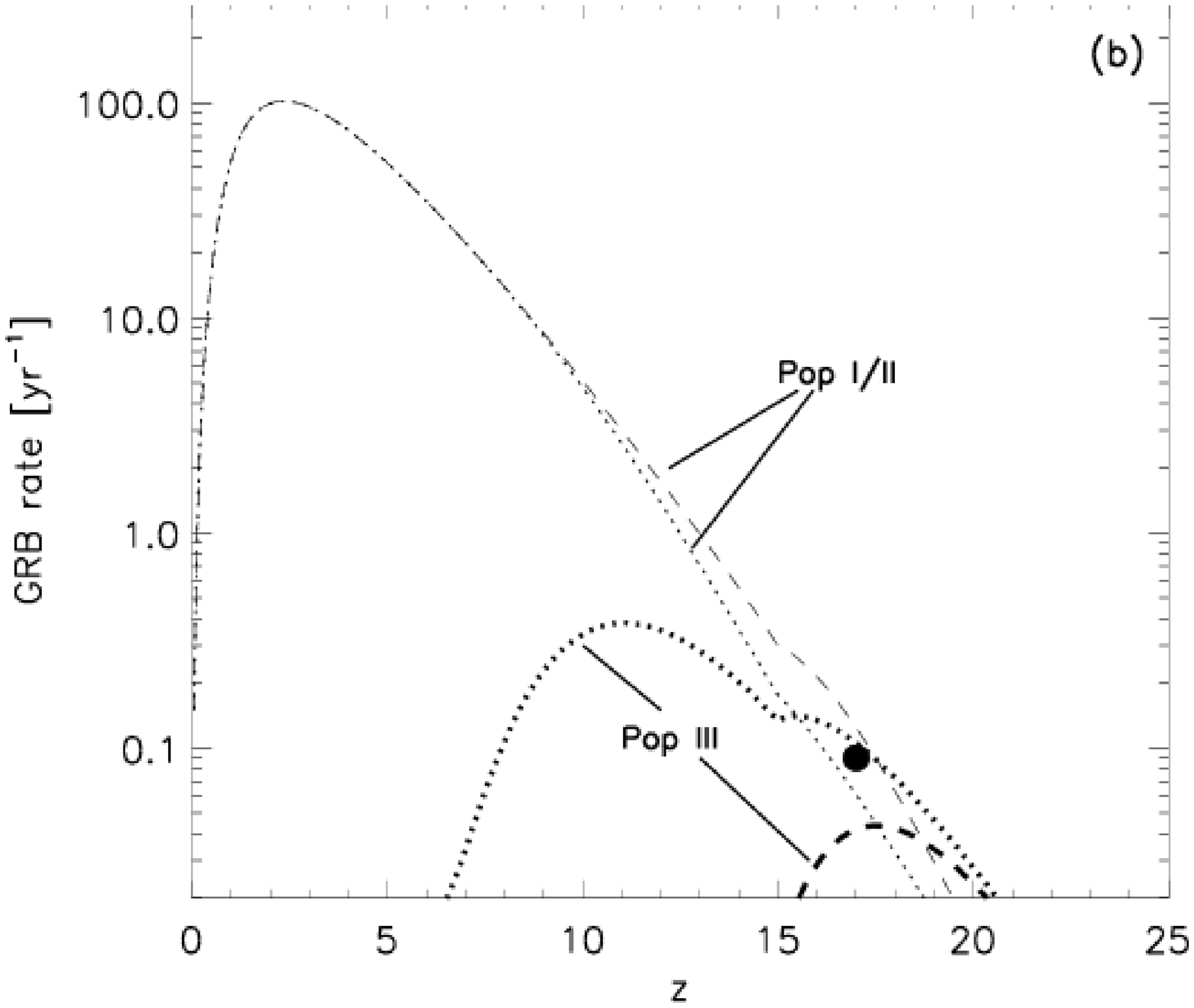}
\end{center}
\vspace{-0.5cm}
\caption{Predicted GRB rate to be observed by {\it Swift} (from Bromm \&
Loeb 2006).  Shown is the observed number of bursts per year, $dN_{\rm
GRB}^{\rm obs}/d\ln (1+z)$, as a function of redshift.  All rates are
calculated with a constant GRB efficiency, $\eta_{\rm GRB}\simeq 2\times
10^{-9}$~bursts $M_{\odot}^{-1}$, using the cosmic SFRs from the previous
figure.  {\it Dotted lines:} Contribution to the observed GRB rate from
Pop~I/II and Pop~III for the case of weak chemical feedback.  {\it Dashed
lines:} Contribution to the GRB rate from Pop~I/II and Pop~III for the case
of strong chemical feedback.  {\it Filled circle:} GRB rate from Pop~III
stars if these were responsible for reionizing the universe at $z\sim 17$.
\label{fig9}
}
\end{figure}

The cosmic star formation history at high redshifts is of great interest
for our origins, since it started filling the universe with heavy elements
such as carbon, oxygen or iron from which rocky planets and eventually
living organisms are made.  The star formation rate (SFR), together with
possible variations in the stellar initial mass function (IMF), affect
models of the initial stages of reionization (e.g., Wyithe \& Loeb 2003;
Ciardi, Ferrara, \& White 2003; Sokasian et al. 2004; Yoshida et al. 2004;
Greif \& Bromm 2006; Johnson et al. 2007) and metal enrichment (e.g.,
Mackey et al. 2003; Furlanetto \& Loeb 2003, 2005; Scannapieco et al. 2003;
Schaye et al. 2003; Simcoe, Sargent, \& Rauch 2004). Due to the
inhomogeneous distribution of metals in the early universe, different modes
of star formation occur simultaneously: massive Pop~III stars in regions of
pristine gas, and normal mass Pop~I/II stars in already enriched
pockets. High-$z$ GRBs will trace the overall cosmic SFR, regardless of
whether the GRB progenitor belongs to Pop~III or Pop~I/II. Furthermore, it
might be possible to separately probe Pop~III star formation, provided that
metal-free stars can trigger a GRB explosion with a signature that is
clearly distinguished from the general population observed at lower
redshifts (see Section 15.2).  Constraints on the Pop~III SFR from ongoing and
future high-$z$ GRB redshift surveys will be extremely valuable to
determine whether planned missions, such as the {\it James Webb Space
Telescope (JWST)}, will be able to effectively probe the first stars (e.g.,
Scannapieco et al. 2005).  The constraints on Pop~III star formation will
also determine whether the first stars could have contributed a significant
fraction to the cosmic near-IR background (e.g., Santos, Bromm, \&
Kamionkowski 2002; Salvaterra \& Ferrara 2003; Kashlinsky et al. 2005;
Madau \& Silk 2005; Dwek, Arendt, \& Krennrich 2005).

Theoretical models of the high-redshift cosmic star formation history have
a number of free parameters, such as the star formation efficiency and the
strength of the chemical feedback. The latter refers to the timescale for,
and spatial extent of, the distribution of the first heavy elements that
were produced inside of Pop~III stars and subsequently dispersed into the
IGM by SN blast waves. From these SFRs, one can derive theoretical GRB
redshift distributions, and one can then use the GRB redshift distribution
observed by {\it Swift}, and any future missions such as EXIST, to
calibrate the free model parameters. This technique could allow us to
measure the redshift where Pop~III star formation terminates, which in turn
is a key ingredient in the modelling of reionization.  In Figure~15.8 and 15.9,
we illustrate this approach (based on Bromm \& Loeb 2006).  Figure~15.9 leads
to the robust expectation that $\sim 10$\% of all {\it Swift} bursts should
originate at $z > 5$. This prediction is based on the contribution from
Population~I/II stars which are known to exist even at these high
redshifts. Additional GRBs could be triggered by Pop~III stars, with a
highly uncertain efficiency (see the discussion in 15.2).

A key ingredient in determining the underlying star formation history from
the observed GRB redshift distribution is the GRB luminosity function,
which is only poorly constrained at present.  The improved statistics
provided by {\it Swift} will enable the construction of an empirical
luminosity function (see the preliminary studies by Salvaterra \&
Chincarini 2007; Guetta \& Piran 2007). With an improved luminosity
function and a better understanding of the observational selection effects
for redshift measurements in GRB afterglows, one will be able to
re-calibrate the theoretical prediction in Figure~15.9 more reliably.

Recent observations have indicated that long-duration GRBs preferentially
originate in regions of low metallicity (Fruchter et al. 2006).  The
evidence is clear for low redshifts, $z< 0.25$, and low-luminosity GRBs
(Stanek et al. 2006). However, at higher redshifts and higher luminosities,
the GRB environments appear to have metallicities that are somewhat {\it
higher} than found in damped Ly$\alpha$ systems of similar hydrogen column
densities (Savaglio, Fall, \& Fiore 2003; Savaglio \& Fall 2004;
see Fig. 3 in Prochaska et al. 2007).
Since stars are expected
to form in the dense gaseous environments traced by damped Ly$\alpha$ at all
redshifts, it appears that the progenitor stars of luminous GRBs do not
originate preferrentially in systems with exceedingly low-metalicities
(for a recent review, see Savaglio 2006). The
metallicity dependence of GRB progenitors has dramatic implications for the
expected number counts of GRBs at high redshifts $z> 5$ (e.g., Salvaterra et
al. 2007) and should be studied further.\\

\noindent {\bf Acknowledgements}

\noindent We acknowledge support from NASA {\it Swift}
grant NNX07AJ636.

\end{document}